\documentclass{iau_FM}
\usepackage{graphicx}
\usepackage{natbib}

\title[Looking into the world of interacting supernovae] 
{Looking into the world of interacting supernovae}

\author[Anjasha Gangopadhyay]   
{Anjasha Gangopadhyay}

\affiliation{Oskar Klein Centre, Department of Astronomy, Stockholm University, AlbaNova, SE-106 91 Stockholm, Sweden \\ email: {\tt anjashagangopadhyay@gmail.com} \\[\affilskip]
}

\pubyear{2024}
\jname{Astronomy in Focus, Focus Meeting 4} 
\editors{Diana M.~Worrall, ed.}
\begin{document}

\maketitle

\begin{abstract}
Interacting supernovae provide key insights into the mass-loss processes of massive stars and their circumstellar environments. By analyzing their photometric and spectroscopic properties, we can study the complex interactions between ejected material and circumstellar material (CSM). This paper highlights the diversity of interacting SNe, including Types IIn, Ibn, and Icn, and explores the challenges in understanding progenitor systems, CSM structures, and late-time evolution. Advances in high-cadence observations and modeling are crucial for improving our knowledge of these stellar explosions.
\end{abstract}

\firstsection 
\section{Introduction to Interacting Supernovae \& Mass-Loss Mechanism}

Recent advancements in wide-field synoptic surveys have led to the detection of nearly 10,000 astronomical transients by 2024, most brighter than 19 mag ($V$-band). Among these, around 4,000 have been identified as supernovae (SNe), though many remain unclassified due to a lack of follow-up observations\footnote{https://sites.astro.caltech.edu/ztf/bts/bts.php}. This influx of data has uncovered rare phenomena, particularly SNe interacting with circumstellar material (CSM). In such dense environments, collisions between ejected material and CSM produce shocks that emit high-energy radiation, shedding light on the complexity of supernovae. Progenitor stars may become unstable before explosion, a process not fully captured by stellar evolution models, and binary interactions likely contribute to the non-spherical nature of CSM interaction.

While supernova diversity was once thought to stem mainly from progenitor properties like mass and rotation, recent studies show that the surrounding environment, especially CSM, plays a significant role. High-cadence surveys reveal that around 10$\%$ of core-collapse SNe exhibit CSM interaction \citep{Perley2020},  offering insights into pre-explosion evolution. Spectral studies and simulations now explore a range of interactions, from hydrogen-rich to hydrogen-free cases, broadening our understanding of stellar explosions.

Mass loss in massive stars, which drives CSM interaction, is governed by stellar luminosity and metallicity \citep{Smith2014}, though research shows that winds are often clumpy, affecting loss rates \citep{2006Fullerton}. Episodic eruptions, like $\eta$ Carinae’s Great Eruption, also play a role, possibly driven by explosive events or binary interactions \citep{2003Smith}. Binary interactions significantly shape mass loss, with around 70$\%$ of massive stars in binaries \citep{2011Sana}.

\citet{1990Schlegel} introduced the IIn classification for SNe with narrow spectral lines. Since then, about 100 Type IIn SNe, characterized by early CSM interaction, have been documented. Related types include Ibn SNe with helium-rich CSM \citep{2007Pastorello}, Icn SNe with carbon/oxygen-rich CSM \citep{2022Pellegrino}, and the rare Ien SNe with nitrogen-rich CSM \citep{2024Schulze}, each highlighting the diversity of SN-CSM interactions.

\section{Photometric and Spectral properties}

Photometry and spectroscopy are key tools for studying SNe. Photometry reveals the light curve behavior and energy distribution, while spectroscopy provides information on the composition and velocity of the ejecta. In SNe interacting with CSM, this interaction slows the ejecta, converting kinetic energy into radiation and boosting luminosity, as seen in Type IIn and Ibn SNe. The CSM reprocesses the radiation, producing narrow spectral lines, such as H$\alpha$ in hydrogen-rich CSM or He lines in helium-rich CSM. Pre-explosion variability suggests episodic mass loss, though connecting progenitors to these SNe remains challenging due to uncertainties in the mass loss mechanisms and pre-explosion phases.

\subsection{SNe~IIn}
\begin{figure}[ht]
        \centering
        \includegraphics[width=0.435\textwidth]{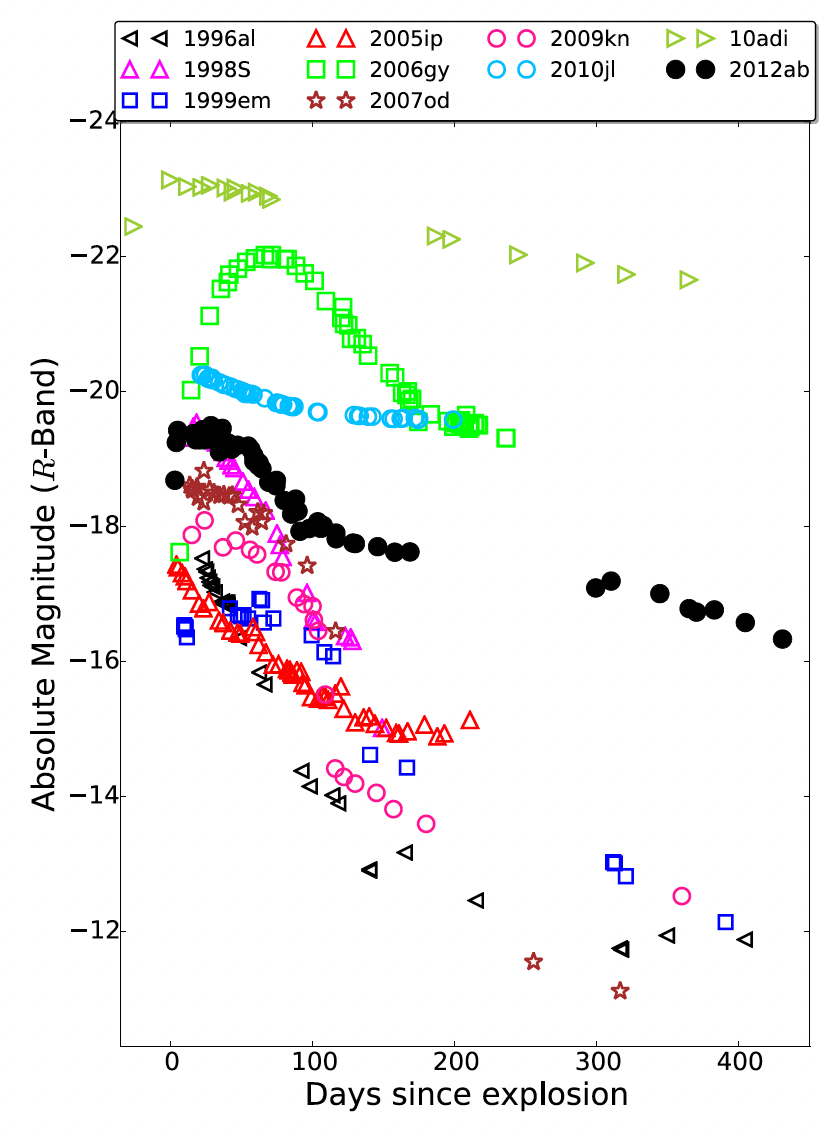}  
         \includegraphics[width=0.45\textwidth]{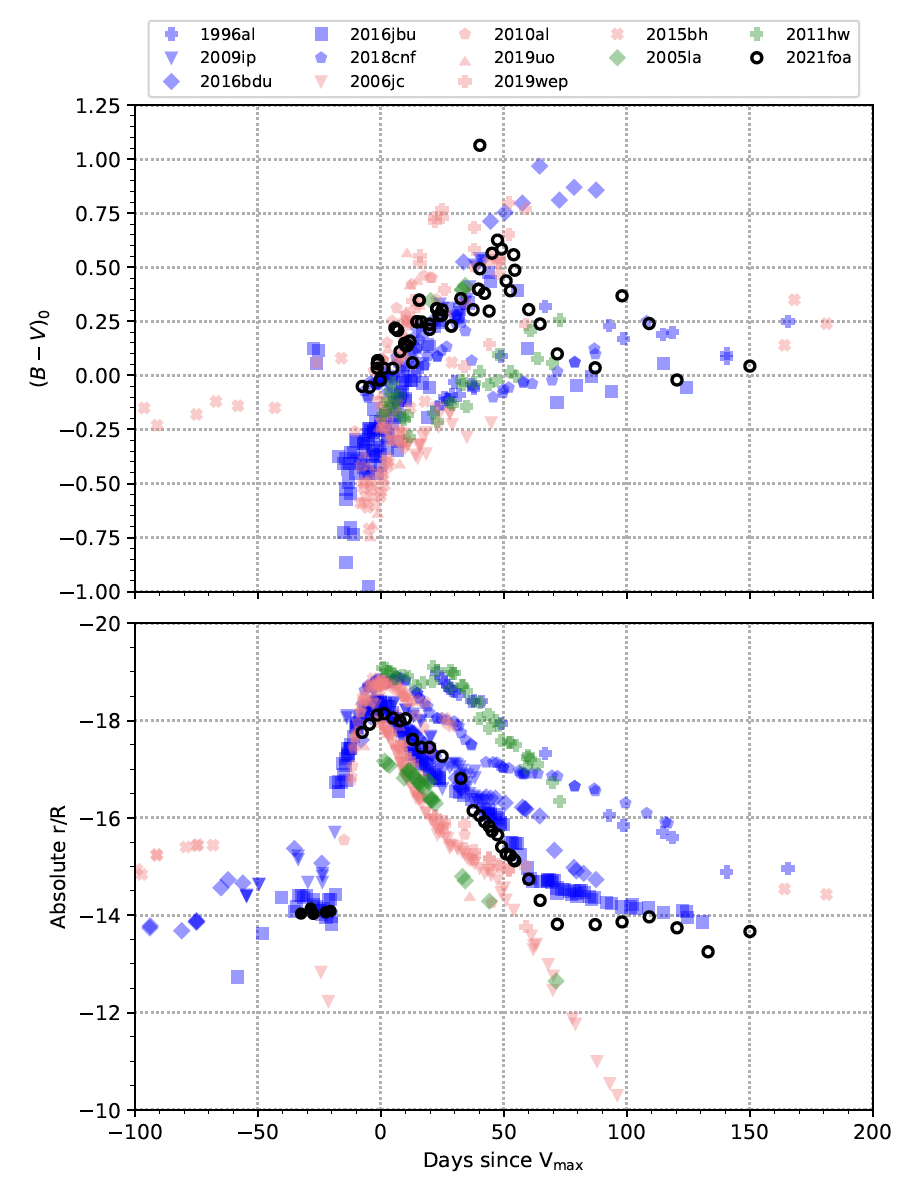}
        \caption{Figure 1 (left) shows the absolute magnitude lightcurve evolution of a group of SNe~IIn which shows examples of long-lived, short-lived and SNe IIn-P.  Figure 1 (right) shows the absolute magnitude and color curves of a group of 2009ip like SNe~IIn and a sample of SNe Ibn see \citep{2020ApJ...889..170G,2020Gangopadhyay}.} 
        \label{fig:lightcurve}
    \end{figure}

There is a huge diversity in the lightcurve and spectral evolution of SN~IIn. A number of Type IIn SNe display slow evolving, long-lived light curves \citep{1991Stathakis}, and are commonly believed to be associated with the explosion of a very massive star of a clearly affected by CSM interaction \citep{Goodrich1989}. The archetype of this group is SN 2010jl, where combined X-ray, radio, and optical observations allowed for detailed modelling of the CSM density \citep{2016Jencson}. Other examples of long-lived interaction in Type IIn SNe include cases of SNe 2012ab, ASASSN-14il \citep{2020Gangopadhyay,2024Dukiya}. The diversity among the SNe~IIn lightcurves are huge, with SNe~IIn-P whose lightcurves show pleatue like behaviour while spectral signatures are like a traditional SNe~IIn. Typical examples include SNe~1994W, 2009kn which have very low Ni masses in the order of 10$^{-2}$ - 10$^{-3}$ M$_{\odot}$   \citep{2013MNRAS.431.2599M}.  Also lies in this zoo are the short-lived interacting members like SN 1998S, whose lightcurve behaved like a SN~IIL, but spectroscopically show narrow emission lines of Hydrogen. Such cases are also seen for SNe~IIPs with a compact dense CSM around eg. SN~2023ixf \citep{2024arXiv240520989S}. There are a number of interacting SNe which show precursor activities associated with the progenitors. They usually also show multi-staged lightcurve behaviours due to different densities and geometries of the CSM. A detailed view of all these sub-classes of diverse asymmetric SNe~IIns are shown in Figure~\ref{fig:lightcurve}.
    
SNe IIn exhibit distinctive spectroscopic features, primarily strong, narrow emission lines superimposed on broader components. These features arise from the interaction between the SN ejecta and a dense CSM, which is often associated with pre-supernova mass loss. Studies have highlighted that asymmetries in the line profiles can be attributed to anisotropic mass loss and the irregular distribution of the CSM, likely due to episodic eruptions of the progenitor star before explosion . This complex CSM environment plays a key role in shaping the observed spectra, with the narrow lines providing insight into the progenitor's mass-loss history and the broad wings indicating interaction with faster-moving ejecta. For instance, \citep{2000ApJ...536..239L} discuss how asymmetric profiles in SN 1998S suggest a clumpy or asymmetric CSM, while \citep{2009ApJ...697L..49S}  emphasize the importance of pre-explosion eruptions in massive stars like LBVs, which lead to dense shells and winds that shape the spectra of SNe IIn . \citep{2020Gangopadhyay,2024Dukiya} also show the existence of a disk like CSM (See Figure~\ref{fig:spectra}). These asymmetries are crucial for understanding the nature of the progenitors and their evolution prior to the SN event. 

\subsection{SNe Ibn and Icn}

SNe Ibn and Icn are relatively new classes of interacting SNe characterized by narrow helium and carbon/nitrogen lines. Type Ibn SNe display rapid light curve evolution, with a fast rise and swift decline, driven by interaction with He-rich CSM \citep{2017ApJ...836..158H}. While SNe Ibn show some diversity in peak magnitudes and decline rates, they are more homogeneous than SNe IIn (see Figure~\ref{fig:lightcurve}). In contrast, the rarer SNe Icn exhibit even faster photometric evolution with shorter-lived light curves \citep{2022ApJ...938...73P}. Lightcurve modeling for both types suggests CSM shell/wind masses of less than 1 M${\odot}$ and nickel masses below 10$^{-2}$ M${\odot}$ \citep{2022ApJ...938...73P}.

\begin{figure}[ht]
        \centering
        \includegraphics[width=0.40\textwidth]{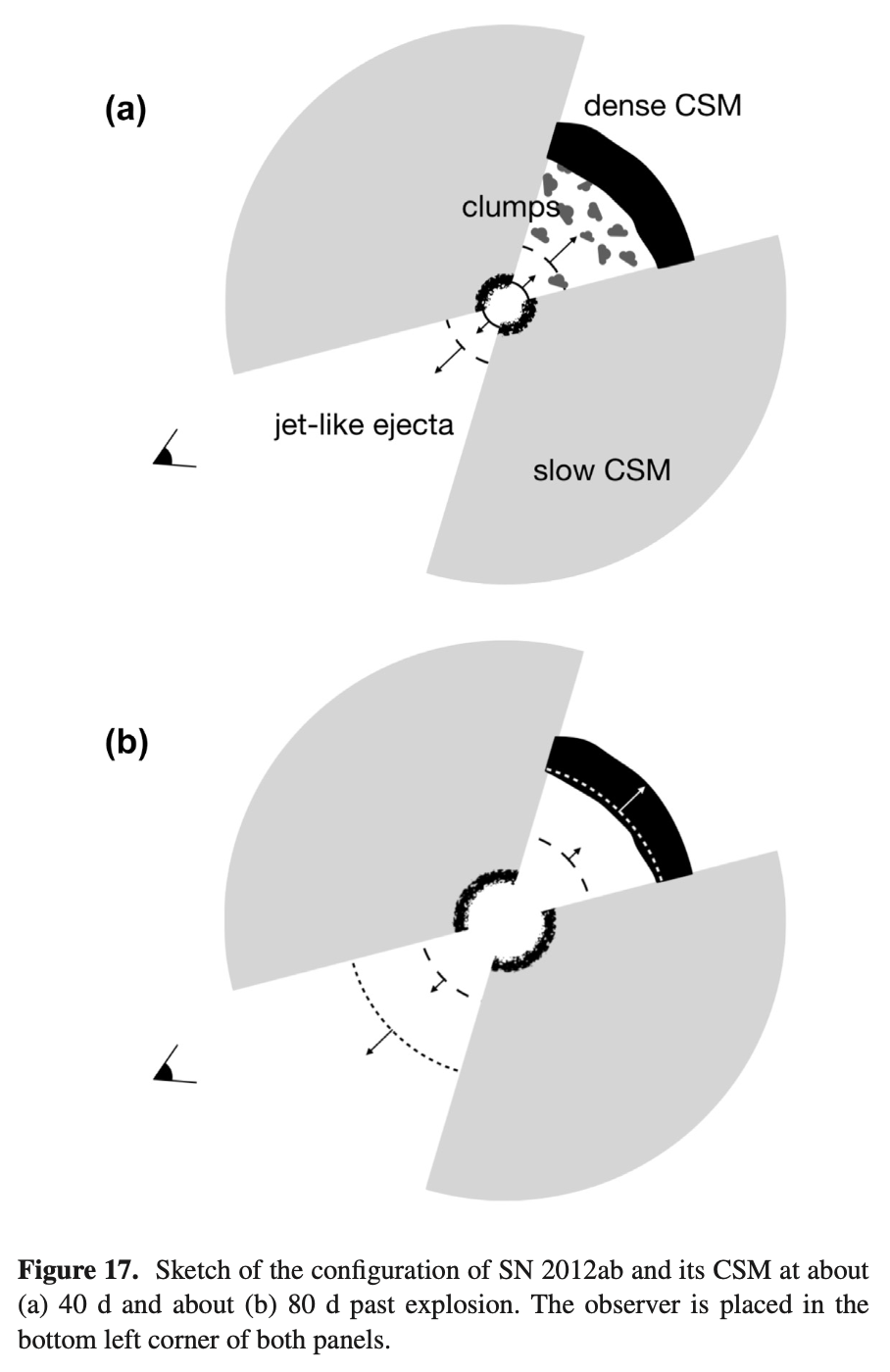}  
         \includegraphics[width=0.43\textwidth]{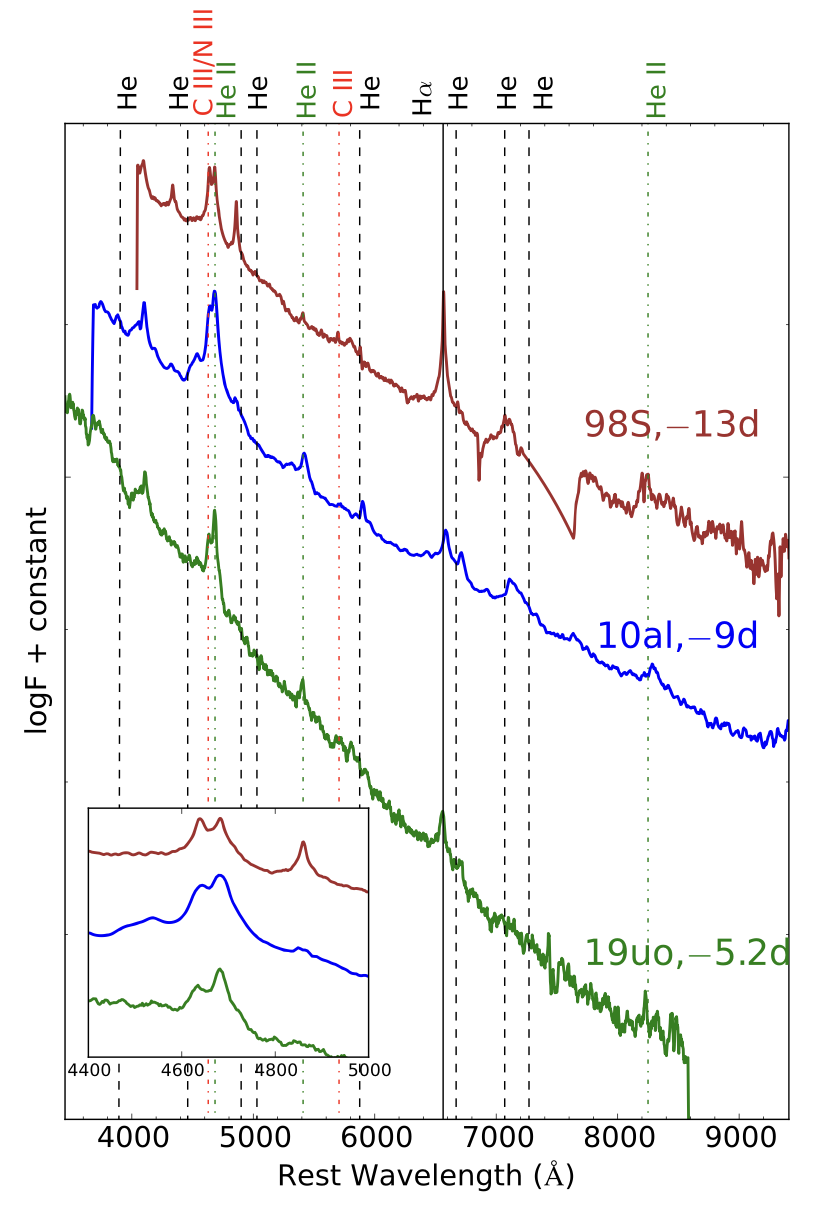}
        \caption{Figure (left) shows the disk like CSM structure for a group of interacting SNe. The H$\alpha$ profile shows complex asymmetric structure with a disk like geometry.  Figure (right) shows the flash ionisation spectral signatures of a group of SNe~Ibn, IIn due to recombination with a dense CSM see \citep{2020ApJ...889..170G,2020Gangopadhyay}.}
        \label{fig:spectra}
    \end{figure}

Spectroscopically, these SNe show narrow He and carbon/nitrogen lines, with early flash ionization features detected in SNe Ibn, indicating Wolf-Rayet progenitors \citep{2020ApJ...889..170G}. The narrow He lines suggest mass loss from pulsational instabilities or binary interactions in WC or WN stars \citep{2020ApJ...889..170G} (see Figure~\ref{fig:spectra}).

\subsection{Late-time interaction and transitioning SNe}
Some SNe, initially classified as one type, can transition to another as interaction signatures become more evident over time. A prime example is SN 2004dj, a typical Type IIP SN during its early phases, but later developed narrow hydrogen emission lines due to interaction with CSM, suggesting that it had experienced significant mass loss prior to explosion \citep{2007ApJ...659.1220C}. Another well-known case is SN 2014C, which transitioned from a hydrogen-poor Type Ib SN to displaying strong hydrogen emission lines typical of a Type IIn, indicating delayed interaction with a dense, hydrogen-rich CSM \citep{2017ApJ...835..140M}. These examples highlight the dynamic nature of SNe and how late-time CSM interaction can drastically alter their classification, shedding light on the complex environments surrounding massive stars before their demise. \citet{2024arXiv240902666G} also find the case of a transitioning SN 2021foa which showed narrow lines of Hydrogen early on and later showed narrow emission lines of Helium, transitioning the class from SNe IIn to SNe Ibn. 

\section{Discussion: Challenges in the Study of Interacting Supernovae}
Interacting supernovae, such as Types IIn, Ibn, and Icn, provide insights into the mass-loss processes of massive stars, but pose several challenges. A key issue is the varied nature of circumstellar material (CSM), which can differ in composition and structure. Understanding the origins—whether from episodic outbursts, clumpy winds, or binary interactions—remains an open question.

Another challenge is identifying progenitors and the mechanisms driving pre-explosion mass loss. Transitions in supernova types, like SN 2014C and SN 2021foa, complicate classification and highlight the need for deeper understanding of late-time CSM interaction.

Limited early-stage observations of interacting SNe further hinder progress. High-cadence, multi-wavelength data are essential to capture the dynamics of CSM interaction, while complex radiative and hydrodynamic models require precise observations for validation.

In summary, while interacting SNe have expanded our understanding of stellar evolution, key questions about progenitors and CSM origins remain, requiring continued observations and refined models.

\end{document}